\documentstyle[multicol,prl,aps,epsfig]{revtex}

\begin{document}
\title{ Halogen-mediated exchange in the coupled-tetrahedra quantum spin systems
       $\rm Cu_2Te_2O_5X_2$ (X=Br,Cl)
         }

\author{Roser Valent\'\i$^1$, T. Saha-Dasgupta$^2$, Claudius Gros$^1$
            and H. Rosner$^{3,*}$}

\address{$^1$Fakult\"at 7, Theoretische Physik,
         University of the Saarland,
         D-66041 Saarbr\"ucken, Germany.}

\address{$^2$S.N. Bose National Centre for Basic Sciences, JD Block, Sector 3,
          Salt Lake City, Kolkata 700098, India.}

\address{$^3$ Department of Physics, University of California, Davis CA
95616}

\date{\today}
\maketitle

\begin{abstract}
Motivated by recent discussion on possible quantum critical behavior
in the coupled Cu-tetrahedra system $\rm Cu_2Te_2O_5Br_2$,
 we present a comparative {\it ab initio} study of the electronic properties
 of 
$\rm Cu_2Te_2O_5Br_2$ and the isostructural $\rm Cu_2Te_2O_5Cl_2$. A detailed
 investigation of the copper-copper interaction pathes reveals that
the halogen-ions play an important role in the inter-tetrahedral
couplings via X$_4$-rings (X=Br, Cl). We find
that, contrary to initial indications,
both systems show a similar electronic behavior with long
range exchange pathes mediated by the X$_4$-rings.

\end{abstract}


%%%%%%%%%%%%%%%%%%%%%%%%%%%%%%%%%%%%%%%%%%%%%%%%%%%%%%%%%%%%%%%%%%%
%%%%%%%%%%%%%%%%%%%%%%%%%%%%%%%%%%%%%%%%%%%%%%%%%%%%%%%%%%%%%%%%%%%

\begin{multicols}{2}

The recently discovered
\cite{johnsson00} spin-tetrahedral compounds $\rm Cu_2Te_2O_5X_2$ (X=Cl, Br)
open the possibility to study the interplay  between localized
many-body tetrahedral cluster excitations and inter-tetrahedral
magnetic couplings leading to a quantum-phase transition with
various possible ordered states \cite{brenig01}. Transitions to
ordered states have been observed experimentally \cite{lemmens02} in
these compounds
with $T_N^{(\rm Br)}=11.4~{\rm K}$ and $T_N^{(\rm Cl)}=18.2~{\rm K}$.
These phase-transitions
exhibit unusual magnetic-field dependences which 
have been linked to the closeness to a quantum-critical point \cite{lemmens02,gros02}.
Unconventional Raman-scattering has been found in the
magnetic channel \cite{brenig01,lemmens02} and the occurrence of low-lying
singlet excitations has been proposed \cite{johnsson00} and observed
by Raman together with a longitudinal magnon \cite{gros02}.

The nature of the ordered states in $\rm Cu_2Te_2O_5X_2$ has not yet
been definitely settled. There is some  evidence for a N\'eel state 
from thermodynamic and susceptibility experiments
and  $\rm Cu_2Te_2O_5Br_2$ has been
proposed to be closer to a non-magnetic singlet state than
$\rm Cu_2Te_2O_5Cl_2$ \cite{gros02}. Thus, it is 
 important
to examine the microscopic behavior of both $\rm Cu_2Te_2O_5Br_2$
and $\rm Cu_2Te_2O_5Cl_2$ by {\it ab-initio} methods and investigate
whether the electronic properties reveal some non-trivial differences
between  the two systems.  The
magnetic exchange coupling parameters in these systems estimated
from susceptibility measurements\cite{johnsson00} are small, of the
order  $40-50~\mbox{K}$, implying a 
small bandwith for the electronically active Cu-3$d$ orbitals close
to the Fermi-edge. 

Here we present a comprehensive first-principles density functional
theory (DFT) study of the electronic properties of these systems
 within the local spin density (LSDA) and the generalized gradient 
approximation (GGA)\cite{Perdew_96}.

{\bf Crystal structure}.- Both  $\rm Cu_2Te_2O_5Br_2$ and  $\rm
Cu_2Te_2O_5Cl_2$ systems  crystallize in the non-centrosymmetric 
tetragonal 
 $P\bar{4}$ space group with two formula units per unit cell.
 $\rm Cu_2Te_2O_5Br_2$ with lattice parameters $a$=7.84 \AA,
$c$=6.38 \AA\ has a larger unit cell than  $\rm Cu_2Te_2O_5Cl_2$ with
 $a$=7.62 \AA, $c$=6.32 \AA.

\begin{figure}[thb]

\centerline{\epsfig{figure=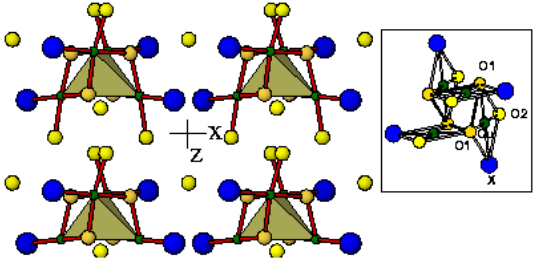,width=10.5cm}}
\caption{ Crystal structure of Cu$_2$Te$_2$O$_5$X$_2$. 
The small green balls in the center of the bonds represent Cu, the
medium sized yellow and large blue balls O and the halogen X,
respectively. The atom not conected to a bond is O3. For 
simplicity the Te atom was droped. The inset
presents an  idealized Cu$_4$-tetrahedron with the four 
corner-sharing O1-O1-O2-X idealized squares and the 
Cu$^{2+}$-ion in the center. 
}
\label{fig_XBS}
\end{figure}

In Fig.~\ref{fig_XBS} we show the crystal structure for $\rm
Cu_2Te_2O_5X_2$. The four equivalent Cu$^{2+}$-ions 
in the unit cell of these systems form distorted
tetrahedra  which are located
on a tetragonal lattice. The structure contains three inequivalent
oxygen positions: The intra-tetrahedral O1, and the inter-tetrahedral
O2  and O3. Of special interest
are the four equivalent halogen-ions, X, per unit-cell which form
together with the O1 and O2 the X-O1-O1-O2 distorted square with the
Cu$^{2+}$-ions in their respective centers, see inset of Fig.~\ref{fig_XBS}.
The inter-tetrahedron distances are slightly smaller in  $\rm Cu_2Te_2O_5Cl_2$ 
than in $\rm Cu_2Te_2O_5Br_2$ due to the size
difference of the res\-pective halogen-ions.
The Cu-Cu intra-tetrahedron distances are,
on the other hand, slightly larger in the Cl-compound.

\begin{figure}[thb]
\centerline{\epsfig{figure=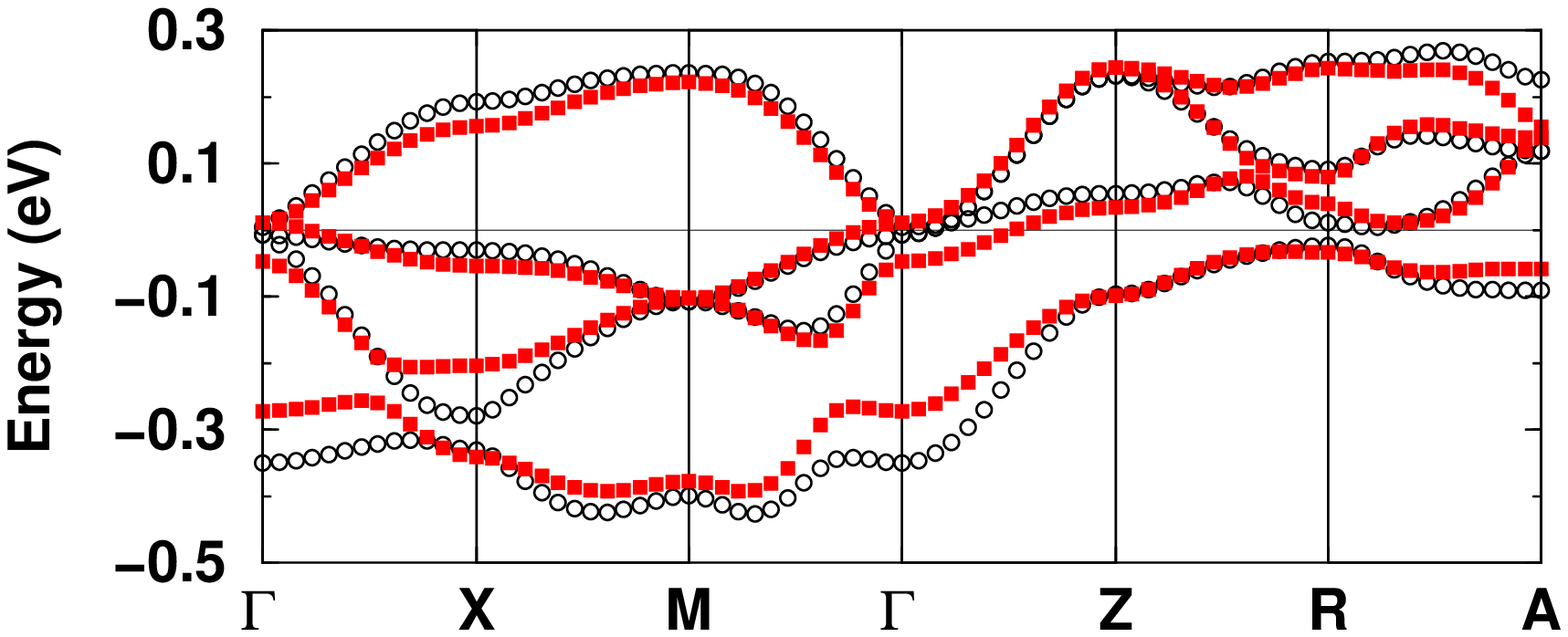,width=9cm}}
         \caption{ Comparison of the band structure near the Fermi
         level between $\rm Cu_2Te_2O_5Br_2$ (full squares)
         and $\rm Cu_2Te_2O_5Cl_2$ (empty circles), with
         $\Gamma=(0,0,0)$, $X=(\pi,0, 0)$, $M=(\pi,\pi,0)$, $Z=(0,0,\pi)$,  
	 $R=(0,\pi,\pi)$ and $A=(\pi,\pi,\pi)$.
%        }
        }
\label{bands_comparison}
\end{figure}

{\bf Band-structure}.- In Fig.~\ref{bands_comparison} we present
 the  non-spin-polarized band-structure of  $\rm Cu_2Te_2O_5Br_2$ 
 and $\rm Cu_2Te_2O_5Cl_2$ near the Fermi-level along various symmetry
 directions
in the Brillouin zone.  Calculations
have been performed  within the framework of the full potential
linearized augmented plane wave (LAPW) method based on the
WIEN97\cite{WIEN97} code, the full potential minimum basis local
orbital code (FPLO)\cite{koepernik99} and the linearized
muffin tin orbital (LMTO) method based on the Stuttgart TBLMTO-47 
code\cite{Andersen_75}.  The band structures obtained by 
the three methods are in overall agreement with each other.

\begin{figure}[thb]
\centerline{\epsfig{figure=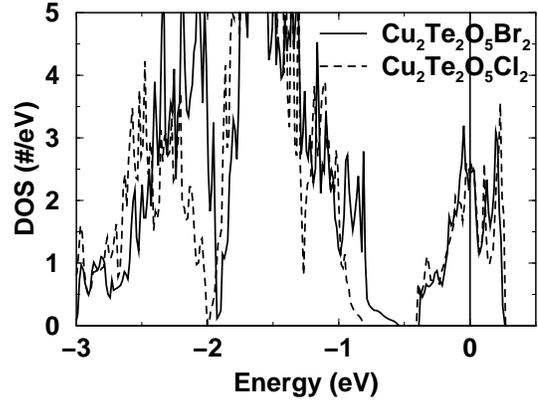,width=7cm}}
\caption{Density  of states for the partial 
         Cu-$3d$ orbitals.
        }
\label{fig_dos_Cu}
\end{figure}

\begin{figure}[thb]
%\centerline{\epsfig{figure=brte_dos_br_O1_O2.eps,width=7.5cm}}
%\centerline{\epsfig{figure=clte_dos_cl_O1_O2.eps,width=7.5cm}}
%\centerline{\epsfig{figure=pDOS_n.eps,width=8cm}}
\centerline{
\epsfig{figure=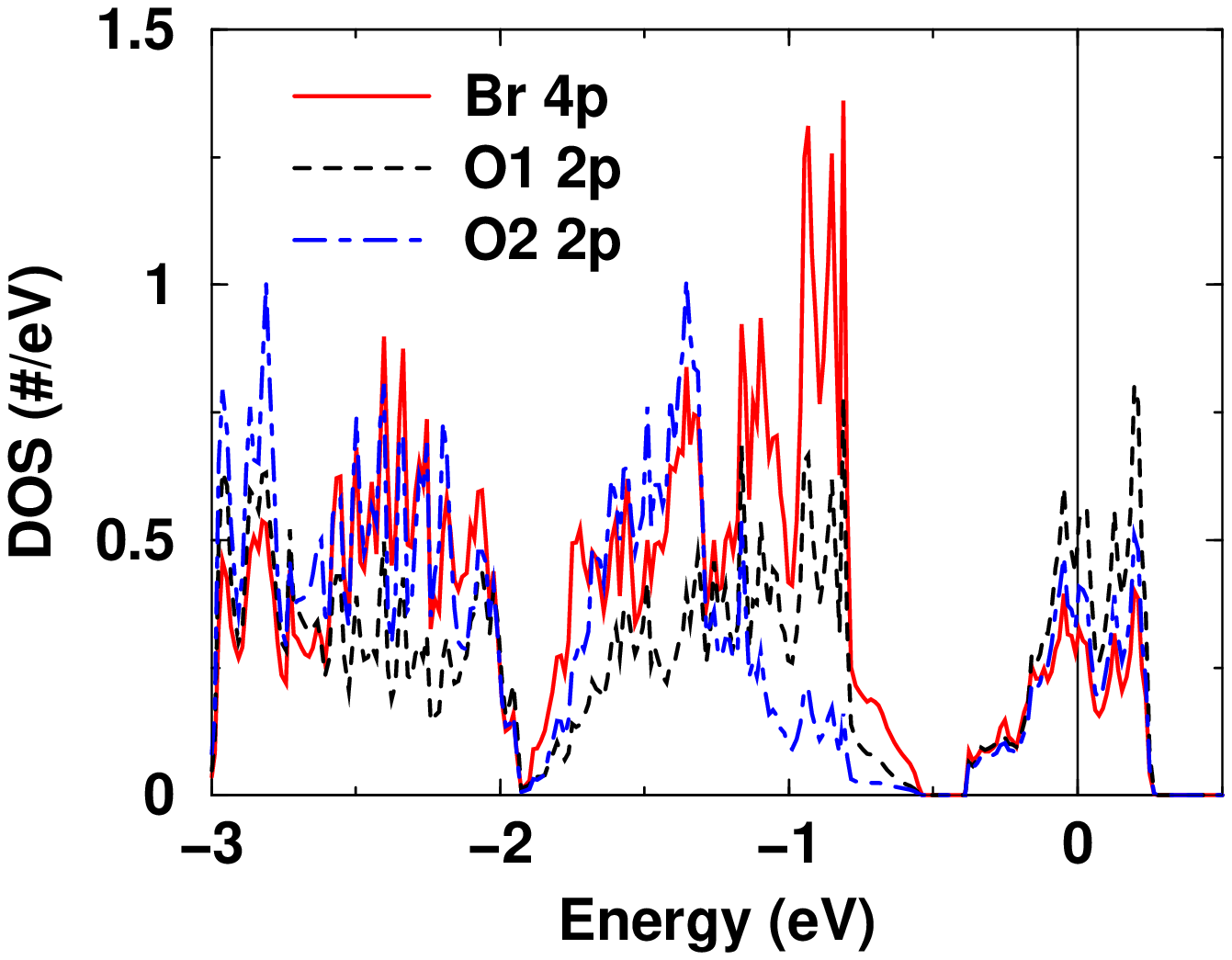,height=3.8cm}
\epsfig{figure=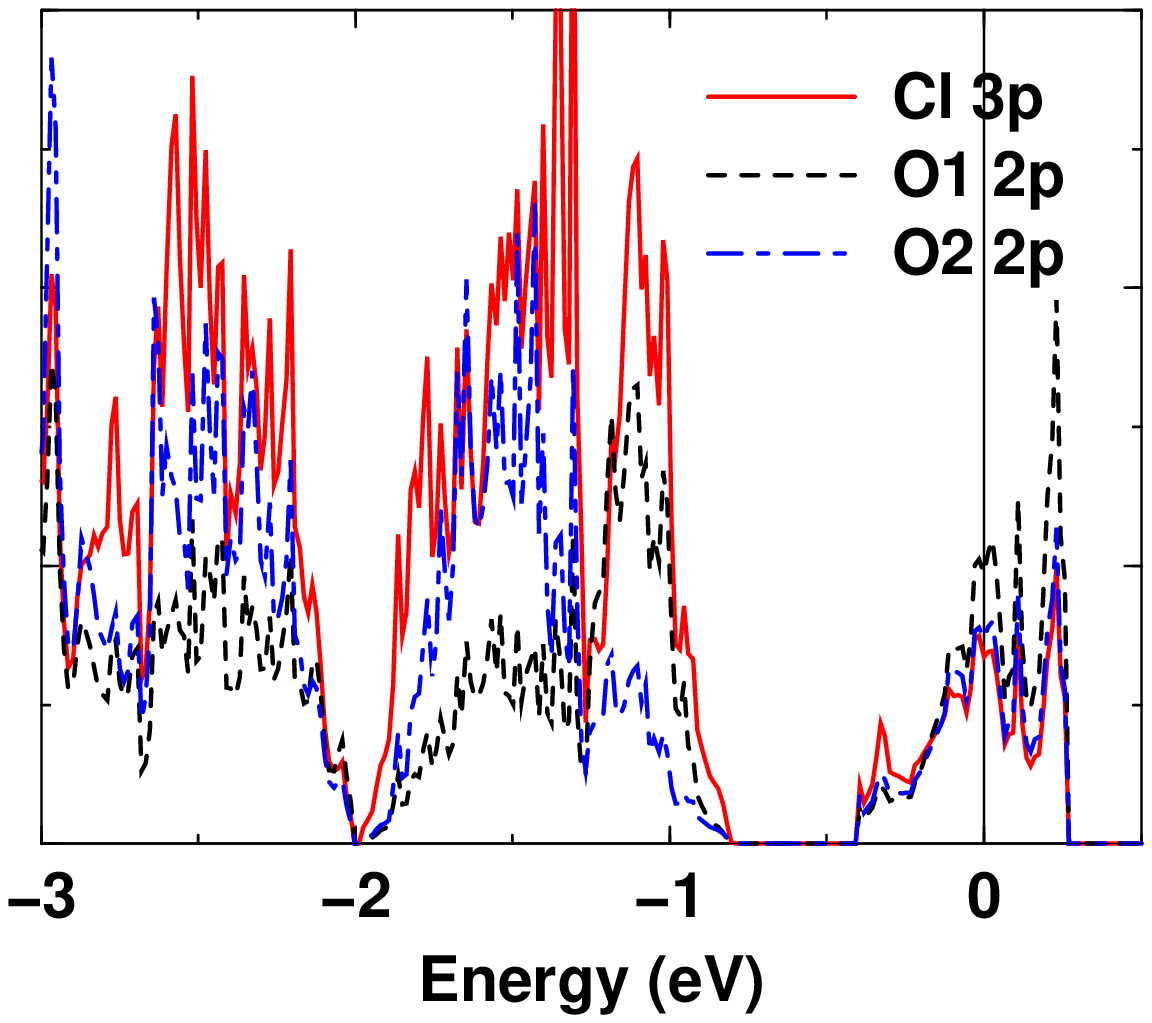,height=3.7cm}
}
\caption{ Density of states for the (total) halogen-$p$ and
         the O1-$p$ and the O2-$p$ orbitals,
	 for $\rm Cu_2Te_2O_5Br_2$ (left) 
	 and $\rm Cu_2Te_2O_5Cl_2$ (right).
        }
\label{fig_dos_XOO}
\end{figure}

The four narrow bands of width $\sim0.7~\mbox{eV}$ 
are well se\-pa\-rated from the occupied  low-lying
valence band by a gap of $\sim0.25~\mbox{eV}$ and the high-lying unoccupied
Te-$p$ bands by a gap of $\sim2~\mbox{eV}$. The bands near the
Fermi-level do not
contain essentially any Te-orbitals.  They are  of dominant Cu-$3d$ character
(predominantly $3d_{x^2-y^2}$ in the local frame of reference) 
as shown in the partial density of
states (DOS) in Fig.~\ref{fig_dos_Cu} with
substantial admixture with Br(Cl) $p$,  O1 and O2 $p$ states 
as shown in Fig.~\ref{fig_dos_XOO}.

\begin{figure}[thb]
\vspace*{-2ex}
\centerline{\epsfig{figure=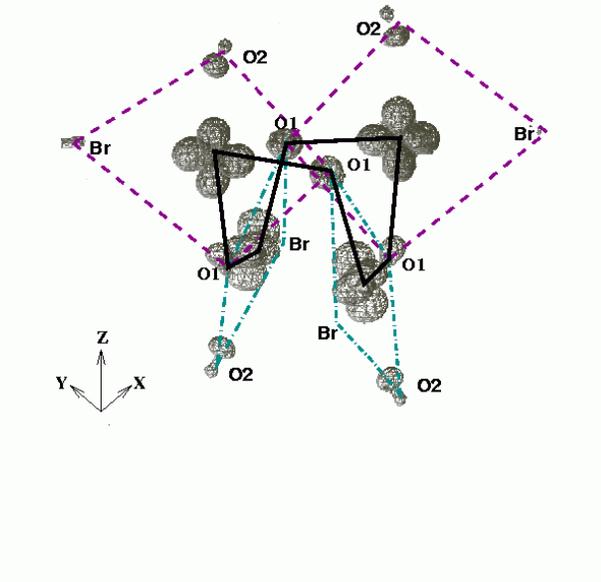,width=8.0cm}
           }
\vspace*{-0ex}
\caption{Electron density for $\rm Cu_2Te_2O_5Br_2$ at the 0.2
         (e/\AA$^3$) isovalue
% Shown is the contribution
%         of the band below the Fermi surface (starting at $\rm -0.3\,eV$)
	 for one tetrahedron. The dashed and da\-shed\--dotted lines
	 indicate the four Br-O1-O1-O2 distorted squares with 
 Cu placed in the respective centers, they share
	 O1-corners. The solid line indicates the Cu-O1-Cu exchange
	 pathes ($t_1$ in  Fig.\ \protect\ref{fig_hoppings}).
        }
\label{fig_ED_tetra}
\end{figure}

 We can investigate these results
further by analizing the electron-density. In Fig.~\ref{fig_ED_tetra}
we present the electron density for $\rm Cu_2Te_2O_5Br_2$ for one
tetrahedral-unit. The lobes of the Cu-$3d$ orbitals are oriented
towards the nearest neighbor Br, O1, O1 and O2 ions, which form the distorted
square surrounding of the Cu ion. 
The covalent bonding in between the Cu-$3d$ orbital and
the respective halogen-$p$ and oxygen-$p$ orbitals leads to the
admixture of halogen and oxygen character at the Fermi-level,
as seen in the respective DOS (Fig.~\ref{fig_dos_XOO}).

Comparing  
the
band-structure 
of both compounds
near the Fermi level (see Fig.~\ref{bands_comparison})
we note that the  bands of both
systems are quite similar and only differ in some details that will
translate in quantitative differences in  the behavior of the effective model.
The band dispersion, due to inter-tetrahedral matrix elements, is
substantial (within the narrow bandwidth) along all three
crystallographic directions indicating 
that the inter-tetrahedral couplings are non-negligible in
all three directions.

%\vspace*{-0.2cm}

  The DFT-calculations yield, within the LDA or
GGA-approximation, to four half-filled (metallic) bands. We performed a spin
polarized calculation which leads to an antiferromagnetic groundstate in
agreement with the experiment, although the resulting band gap is
too small.
An additional local Coulomb-repulsion -- not taken fully into account in
the LSDA or GGA-approximation -- will basically further shift these
bands, enlarging the gap and resulting in the picture of
lower- and upper-Hubbard bands in agreement with
experiment. 
 In the following, however, we did not do the simulation of 
the missing Coulomb interaction in an LDA+$U$ type calculation, which is 
in general useful but we do not expect from it new insights about the 
important couplings.

 Attempts to directly compute the exchange integrals by comparing
the LSDA total energy differences for different spin arrangements to that
of a Heisenberg-like model faced two basic short-comings:
(i) The resulting magnetic moments at the
copper sites were quite different for different spin
arrangements. This leads to a strong bias of the inter-atomic exchange
energies due to different intra-atomic contributions to the total energy.
(ii) The important interactions in these compounds have a rather long-ranged
nature as mentioned below. This would demand the calculation of large
unit cells beyond the present computational capabilities.  
Therefore, 
we focus this letter to accessing the corresponding transfer integrals
thereby pointing out the important interaction path-ways in these systems.

\begin{figure}[thb]
\centerline{\epsfig{figure=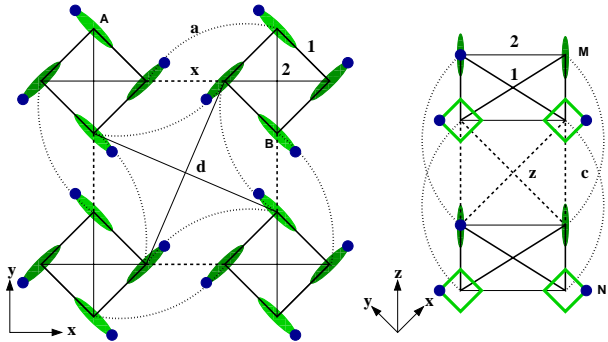,width=8.0cm}}
%\vspace{-0.5cm}
\caption{Illustration of the hopping matrix elements $t_\alpha$
        ($\alpha=1,2,x,z,a,c,d$)
        in between the Cu$^{2+}$ ions  which are located in the center
	of the $\rm O_3X$ distorted square (X=Br,Cl), indicated
	by the shaded regions (not to scale, in reality the distorted squares share O1-corners). 
	The filled circles denote the direction of the
 halogen-sites. The parameters $t_r$ and $t_v$ correspond to the hopping
% matrix 
 elements between Cu$^{2+}$ ions at A and B  and at M and N
 respectively and
 equivalent positions.
        }
\label{fig_hoppings}
\end{figure}

{\bf Effective Model}.- In order to quantify the results obtained from
the {\it ab-initio} calculations in terms of hopping matrix elements, $t_{ij}$,
we have employed LMTO-based downfolding\cite{newlmto} and tight-binding analysis on
the band structure of these systems. The downfolding method consists in
deriving a few-orbital effective hamiltonian from the full  LDA or  
GGA hamiltonian by folding down the inactive orbitals in the tails of the 
active orbitals kept in the basis chosen to describe the low-energy physics 
of the system. This procedure naturally takes into account the proper
renormalization effect of  integrated-out inactive orbitals in the
 effective interactions defined in the basis of the active orbitals. 
Using the real-space description of the downfolded Hamiltonian one gets
$H_R = -\sum_{ij} t_{i,j}\left(
c_j^{\dagger}c_i^{\phantom{\dagger}} + c_i^{\dagger}c_j^{\phantom{\dagger}}
                          \right)$
where $t_{ij}$ provides the effective hopping matrix elements
between the active orbitals.
In Table \ref{table2} we present  the results for the most significant
hopping matrix elements -shown schematically in Fig.\ \ref{fig_hoppings}- obtained from
downfolding 
 the full LMTO Hamiltonian to effective Cu-only Hamiltonian
by integrating out everything except the $d_{x^{2}-y^{2}}$  orbital for
each Cu atom in the unit cell to reproduce accurately the four narrow 
bands close to the Fermi energy.  The first and second column of 
Table~\ref{table2} show the hopping parameters for 
$\rm Cu_2Te_2O_5Br_2$ and $\rm Cu_2Te_2O_5Cl_2$ respectively.

The predominant matrix elements consist of  
a set of nine different hopping parameters, some of them 
quite long-ranged and which couldn't be neglected in order
to get a good description of the energy bands.  Apart from the
intra-tetrahedra $t_1$ and $t_2$ parameters, the
inter-tetrahedra  $t_x$, $t_z$, $t_a$, $t_c$, $t_v$, $t_d$ and $t_r$
are important (see Fig.\ \ref{fig_hoppings}).  
For instance, $t_c$ is needed in order to get dispersion
along the path $\Gamma-M$ and $t_a$ and $t_d$ are essential in order to
get the correct behavior along the path $X-A$.  
The need of including such longer-ranged hoppings like $t_r$ 
is set by the renormalization process of the downfolding procedure\cite{remark_1}.  
We note, that long-ranged hopping matrix elements have proven to be essential 
for the description of some related copper
systems \cite{vt_02}.

In order to investigate more in detail the nature of the Cu-Cu 
interaction paths and in particular the role of the halogen ion, 
we have also performed, a TB-downfolding analysis of the 
bandstructure of both systems by keeping
the halogen X-$p$ orbitals active in addition to four Cu-$d$ orbitals
in the basis set (Cu+X downfolding, see third and fourth column of 
Table \ref{table2}). 
 The results  will be discussed
in the following.

% ---------- %
% ---------- %
\vbox{
\begin{table}[bth]
\caption{TB-Downfolding parameters for the effective Cu-Cu
         hoppings in meV  units obtained by  keeping active (i) the
         Cu-$3d$ orbitals (first and second columns, Cu-downfolding) and
	 (ii) both the Cu-$3d$ with the halogen-$3p/4p$ orbitals 
	  (third and fourth columns, Cu+X downfolding).
        }
\begin{tabular}{|c|c|c||c|c|c|}
& Cu$_{\rm Br}$ & Cu$_{\rm Cl}$ & &Cu+Br & Cu+Cl \\ \hline
$\rm t_1$ & 80 & 98 & $\rm t_1$ & 155 & 181\\
  $\rm t_2$ &  4 & 0 &  $\rm t_2$ &  -156 & -132 \\
$\rm t_x$ & -16 & -10 & $\rm t_x$ & -10 & -14 \\
 $\rm t_z$ & 11 & 12 & $\rm t_z$ & 34 & 33\\
 $\rm t_a$ & -30  & -29 & $\rm t_a$ & 5 & 8\\
 $\rm t_c$ & -48  & -45 & $\rm t_c$ & -26 & -19  \\
  $\rm t_v$ & -24  & -23  & $\rm t_v$ & 9 & 8 \\
  $\rm t_d$ & -73 & -80 & $\rm t_d$ & 8 & 8 \\ 
 $\rm t_r$ & -35  & -48  & $\rm t_r$ & -62 & -72 \\     
\end{tabular}
\label{table2}
\end{table}
}
% ---------- %
% ---------- %

%
{\bf Intra-tetrahedral couplings}.- The nearest neighbor (n.n.) intra-tetrahedral
Cu-Cu transfer matrix element $t_1$ is mediated by 
O1-ions located in between two Cu-ions and responsible
for the Cu-O1-Cu superexchange generating the spin-spin coupling $J_1$.
This superexchange-path is evidenced in Fig.~\ref{fig_ED_tetra}.
The angle Cu-O1-Cu is $107^\circ/109^\circ$ for $\rm X=Br/Cl$.
Assuming a similar crystal field, the larger
superexchange-angle for X=Cl results -following the Goodenough-Kanamori-Anderson
rules\cite{GKA_55}- in $t_1(X=Cl)>t_1(X=Br)$, 
in agreement with the downfolding results, see Table~\ref{table2}. 

The intra-tetrahedral hopping matrix element $t_2$ corresponds to the 
effective next nearest neighbor (n.n.n.) Cu-Cu overlapp. A comparison
of the Cu- and the Cu+X downfolding result 
shows a substantial drop of $t_2$ for both systems when the halogen-orbitals
are integrated out. This drop  indicates that 
pathes of the type Cu-X-X-Cu are important for the effective $t_2$ matrix element 
and that they compensate to a certain extend the contribution of the direct
Cu-Cu path (see Fig.\ \ref{fig_ED_tetra}) and of other possible pathes
through oxygen ions. 
We note, that $t_2$ is vanishingly small in the Cl-compound.

\begin{figure}[thb]
%\centerline{\epsfig{figure=brte_ED_0.05_xy.eps,width=4.1cm}\hfill
%            \epsfig{figure=clte_ED_0.05_xy.eps,width=4.1cm}
%           }
\centerline{\epsfig{figure=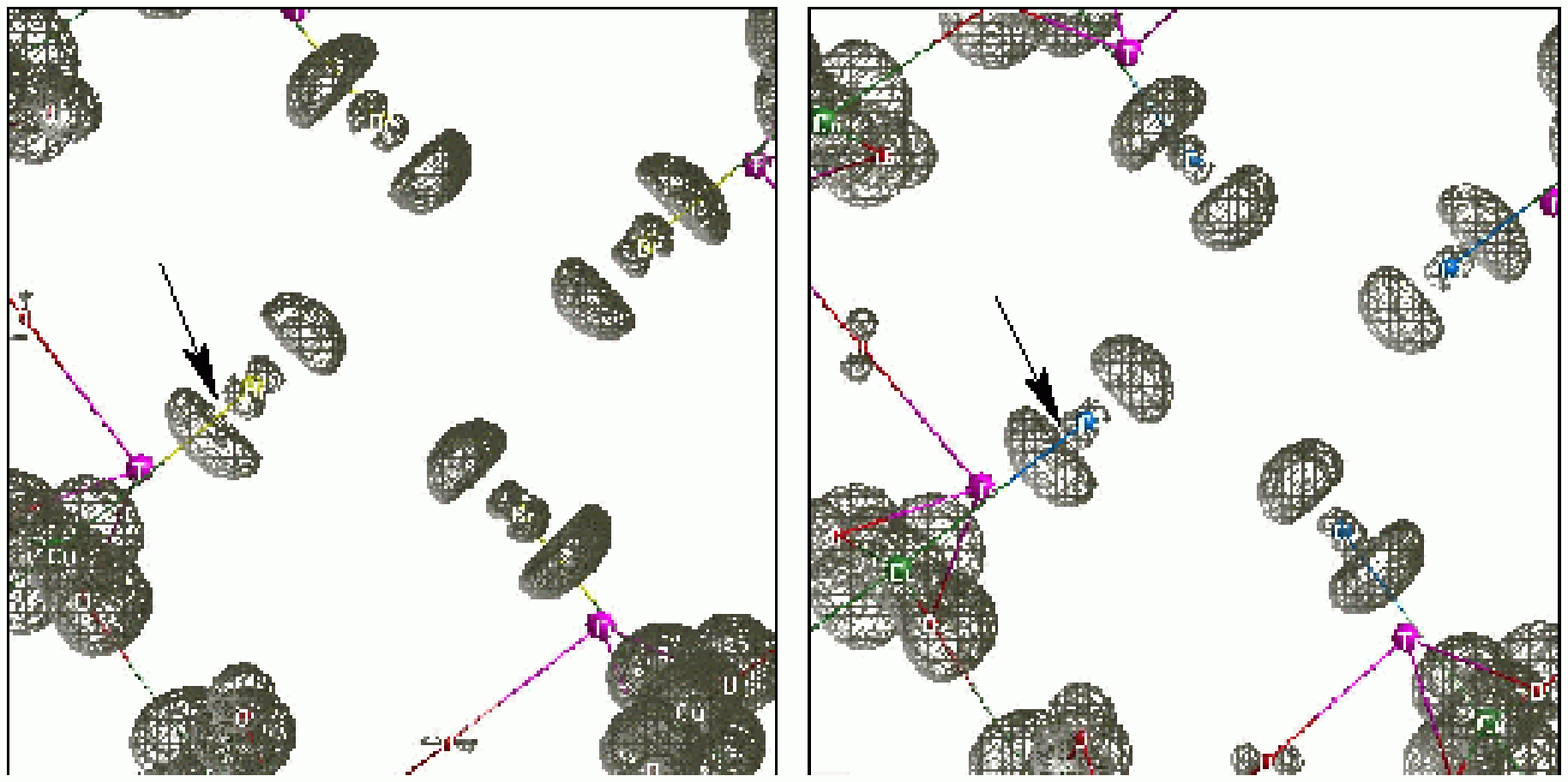,width=9.8cm}}
\caption{Electron density for $\rm Cu_2Te_2O_5Br_2$ (left)
         and $\rm Cu_2Te_2O_5Cl_2$ (right) at the 0.05
	 (e/\AA$^3$) isovalue.
% Shown are the contributions
% of the band below the Fermi surface (starting at $\rm -0.3\,eV$)
	 for a projection on the $xy$-plane in between four
 Cu-tetrahedra (compare with Fig.\ \protect\ref{fig_hoppings}, left).
The four $p$ orbitals at the center are Br-$3p$ (left) and Cl-$4p$
(right). The respective arrows indicate the distortion of the
X-$p$ orbitals.
        }
\label{fig_ED_xy}
\end{figure}

{\bf Inter-tetrahedral coupling}.- 
The coupling $t_x$ via the O3
along the $x-$ and $y-$direction in between two
Cu$_4$-tetrahedra (see Fig.~\ref{fig_hoppings} and 
Table~\ref{table2}), is small, partly
because the O3-weight is small at the Fermi-level.
There is, however, a substantial inter-tetrahedral
coupling $t_a$ and diagonal $t_d$ within the $xy$-plane 
mediated by the halogen-$p$ orbitals, (see Fig.~\ref{fig_hoppings} and 
Fig.~\ref{fig_ED_xy}).  
The role of the halogen ions for the inter-tetrahedral couplings can
be quantified by analysing the Cu-downfolding and the Cu+X-downfolding results presented
in Table~\ref{table2}.  Comparing 
the two corresponding values (Cu$_X$ and Cu+X) we observe that
$t_a$ and $t_d$ are nearly exclusively
due to halogen-containing exchange pathes.

The reason for this unusual large inter-tetrahedron coupling 
in $\rm Cu_2Te_2O_5X_2$
is the large extension of the Cl-3$p$ and Br-4$p$ wave functions 
which do not occur in cuprates containing only
O-2$p$ orbitals (i.e. CuO$_4$ units). In contrast to certain other cuprate compounds 
containing halogen-ions like Sr$_2$CuO$_2$Cl$_2$ \cite{hayn99}, the situation for the
$\rm Cu_2Te_2O_5X_2$-compounds is unique. Here, the halogen is part of
the covalent Cu-O-Cl(Br) network, whereas in the former mentioned
compound family the halogens play only the role of an anionic charge 
reservoir \cite{hayn99}.

A closer analysis of the electron density presented in
Fig.~\ref{fig_ED_xy} leads to two observations, which are
in agreement with Cu+X downfolding results:
(i) The Cl-$3p$ orbital at the Fermi-level is more strongly
distorted towards the Cu-$3d$ orbital  compared to
the Br-$4p$ orbital,
indicating stronger copper-halogen coupling for $\rm X=Cl$
than for $\rm X=Br$ (see arrows in Fig.\ \ref{fig_ED_xy}) 
(ii) The Br-$4p$ orbital is reoriented somewhat towards the
n.n.\ Br-$4p$ orbital indicating a stronger covalent Br-Br overlap.

In addition to the inter-tetrahedral hopping processes
along $x$ and $y$ there are substantial contributions to the
inter-tetrahedral coupling terms $t_c$, $t_v$ and $t_z$
along the $z$-direction, compare Fig.~\ref{fig_hoppings}
and Table~\ref{table2}.

{\bf Discussion}.- 
It has been recently argued \cite{gros02} that
$\rm Cu_2Te_2O_5Br_2$ and $\rm Cu_2Te_2O_5Cl_2$
differ only quantitatively but not qualitatively in their 
magnetic properties, with $\rm Cu_2Te_2O_5Br_2$ being somewhat closer to
a quantum-critical phase transition. Here we present
evidence from {\it ab-initio} calculations that the
electronic properties of these two compounds are indeed close
and that the inter-tetrahedral coupling is considerable.
We note that a subs\-tantial coupling between Cu-tetrahedra
is necessary in order to establish magnetic long-range
order \cite{gros02}.

We have also presented a detailed analysis in terms 
of two different downfolding models (Cu and Cu+X) which
reveal the important result that the halogen-ions are
essential for the inter-tetrahedral exchange couplings.
We find that four halogen orbitals, coupled by considerable
(X-$p$)-(X-$p$) covalent bonding give
rise to X$_4$-rings which mediate long-ranged inter-tetrahedral
couplings. These X$_4$-rings are covalently coupled to the respective
Cu$_4$-tetrahedrons.

{\bf Acknowledgements}.- 
We acknowledge the support of the German 
Science Foundation (DFG SP1073) and
 discussions with F. Mila, P. Lemmens and W. Brenig.
We thank A. Kokalj for providing the graphics XCRYSDEN code.
%%%%%%%%%%%%%%%%%%%%%%%%%%%%%%%%%%%%%%%%%%%%%%%%%%%%%%%%%%%%
%%%%%%%%%%%%%%%%%%%%%%%%%%%%%%%%%%%%%%%%%%%%%%%%%%%%%%%%%%%%

%%%%%%%%%%%%%%%%%%%%%%%%%%%%%%%%%%%%%%%%%%%%%%%%%%%%%
%%%%%%%%%%%%%%%%%%%%%%%%%%%%%%%%%%%%%%%%%%%%%%%%%%%%%%

%\vspace*{-1cm}

% ---------- %
% ---------- %
% ---------- %
% ---------- %

% ---------- %
% ---------- %
% ---------- %
% ---------- %

% ---------- %
% ---------- %
% ---------- %
% ---------- %

%---------- %
%---------- %
% ---------- %
% ---------- %

% ---------- %
% ---------- %
% ---------- %
% ---------- %

% ---------- %
% ---------- %
% ---------- %
% ---------- %

\end{multicols}


\begin{thebibliography}{99}
\bibitem{remark_1}
We would like to mention here that a reduced set of
hopping terms (i.e. $t_1$, $t_2$, $t_a$, $t_d$, $t_z$, $t_x$)
variationally varied to reproduce the behavior of the bands gives
values for $t_2$ comparable to $t_1$ in the Br compound and values of $t_d$
much smaller than those obtained with the larger set of parameters.
This reduced set fails to reproduce accurately the bands-features
along the path $Z-R-A$.
\end{thebibliography}
\end{document}